\documentclass[aps,prc,superscriptaddress,preprint,nofootinbib]{revtex4-1}
\pdfoutput=1

\usepackage{graphicx}
\usepackage{bm}
\usepackage{amssymb,amsmath,latexsym}
\usepackage{color}
\usepackage{xcolor}
\definecolor{darkblue}{RGB}{0,0,196}
\definecolor{darkred}{RGB}{196,0,0}
\usepackage[colorlinks=true,linktocpage=true,linkcolor=darkblue,citecolor=darkblue,urlcolor=darkblue]{hyperref}

\def\be{\begin{equation}}
\def\ee{\end{equation}}
\def\ba{\begin{eqnarray}}
\def\ea{\end{eqnarray}}

\def\xifs{\xi_{\rm FS}}
\def\xifsdot{{\dot \xi}_{\rm FS}}
\def\frs{f_{\rm RS}}

\begin{document}

\title{An improved anisotropic hydrodynamics ansatz}

\author{Huda Alalawi} 
\author{Michael Strickland} 
\affiliation{Department of Physics, Kent State University, Kent, OH 44242, United States}

\begin{abstract}
We introduce an improved form for the anisotropic hydrodynamics distribution function which explicitly takes into account the free-streaming and equilibrating contributions separately.  We demonstrate that with this improvement one can better reproduce exact results available in the literature for the evolution of moments of the distribution function, in particular, for moments which contain no powers of the longitudinal momentum in their definition ($m=0$ moments).  Using the resulting dynamical equations, we extract the non-equilibrium attractor associated with our improved aHydro ansatz and demonstrate that the improvement also allows one to better reproduce the exact dynamical attractor obtained using kinetic theory in the relaxation time approximation, particularly at early rescaled times and for $m=0$ moments.
\end{abstract}

\date{\today}



\maketitle
%\tableofcontents

%%%%%%%%%%%%%%%%%%%%%%%%%%%%%%%%%%%%%%%%%%%%%%%%%%%%%%%%%%%
\section{Introduction}
\label{sec:intro}
%%%%%%%%%%%%%%%%%%%%%%%%%%%%%%%%%%%%%%%%%%%%%%%%%%%%%%%%%%%
 
In the very early universe (a few microseconds after the Big Bang), the quark-gluon plasma (QGP) is believed to have existed where the density can reach values ten times higher than those of ordinary nuclei. It was speculated theoretically that one can reach these extreme conditions by colliding two heavy nuclei with ultrarelativistic energies. In this collision, the temperatures can be million times hotter than the core of the sun, and a fraction of the kinetic energies of the two colliding nuclei transform to heat the QCD vacuum within an extremely small volume. Because of the appearance of modern accelerator facilities, ultrarelativistic heavy-ion collisions (URHICs) be able to provide an opportunity to systematically create and study different phases of the bulk nuclear matter. In heavy-ion collision experiments at Relativistic Heavy-Ion Collider (RHIC) located at Brookhaven National Laboratory, USA, and Large Hadron Collider (LHC) at European Organization for Nuclear Research (CERN), Geneva, the new state of matter (the QGP) is widely believed created. Results obtained at RHIC energies and recently at LHC energies strongly suggested the formation of a quark-gluon plasma (QGP) which may be close to (local) thermodynamic equilibrium, albeit in a tiny volume ($\sim 100-1000\;{\rm fm}^3$). After the QGP is generated, it is expected to expand, cool, and then hadronize in the final stage of its evolution, with a QGP lifetime on the order of 10 fm/c in central collisions~\cite{Averbeck:2015jja,Jeon:2016uym,Romatschke:2017ejr}.  

Heavy-ion collisions such as those at RHIC provide a primary tool to study the thermodynamic and transport properties of the QGP. Of remarkable importance is knowledge of time evolution of the rapidly expanding the QGP that produced in these URHICs.  For this purpose, one can use a basic theoretical approach called relativistic hydrodynamics to describe the QGP.  The resulting models describe the collective behavior of the soft hadrons with $P_T \lesssim 2$ GeV quite well.  In early studies, it was found that the QGP created at RHIC energies was well described by models which assume ideal hydrodynamic behavior from very early times $\tau \lesssim 1$ fm/c \cite{Huovinen_2001,PhysRevC.66.054905,Hirano:2002ds,Kolb:2003dz,kolb2003hydrodynamic}. Traditionally, one can apply ideal hydrodynamics if the system is in perfect isotropic local thermal equilibrium. Based on these early studies, it was expected that the QGP would isotropize on a timescale $\tau \sim 0.5$ fm/c. In practice, however, when one includes viscous corrections to the hydrodynamical models \cite{Muronga:2001zk,Muronga:2003ta,Muronga:2004sf,Heinz:2005bw,Baier:2006um,Romatschke:2007mq,Baier:2007ix,Dusling:2007gi,Luzum:2008cw,Song:2008hj,Heinz:2009xj,El:2009vj,PeraltaRamos:2009kg,PeraltaRamos:2010je,Denicol:2010tr,Denicol:2010xn,Schenke:2010rr,Schenke:2011tv,Bozek:2011wa,Niemi:2011ix,Denicol:2011fa,Niemi:2012ry,Bozek:2012qs,Denicol:2012cn,Denicol:2012es,PeraltaRamos:2012xk,Jaiswal:2013npa,Jaiswal:2013vta,Calzetta:2014hra,Denicol:2014vaa,Denicol:2014mca,Jaiswal:2014isa} one observes that at times  \mbox{$\tau \lesssim 2$ fm/c} there can still be sizable differences between the transverse pressure, ${\cal P}_{T}$, and longitudinal pressure, ${\cal P}_{L}$  which is associated with the existence of a non-equilibrium hydrodynamic attractor~\cite{Chesler:2009cy,Heller:2015dha,Chesler:2016ceu,Keegan:2015avk,Heller:2016rtz,Florkowski:2017olj,Romatschke:2017vte,Bemfica:2017wps,Spalinski:2017mel,Romatschke:2017acs,Behtash:2017wqg,Florkowski:2017jnz,Florkowski:2017ovw,Strickland:2017kux,Kurkela:2018vqr,Kurkela:2018qeb,Kurkela:2019kip,Almaalol:2018ynx,Denicol:2018pak,Behtash:2018moe,Strickland:2018ayk,Heller:2018qvh,Behtash:2019qtk,Strickland:2019hff,Jaiswal:2019cju,Kurkela:2019set,Chattopadhyay:2019jqj,Brewer:2019oha,Heller:2020anv,Almaalol:2020rnu,Blaizot:2020gql,Mitra:2020mei,Dore:2020fiq,Dash:2020zqx,Das:2020gtq,Berges:2020fwq,Kamata:2020mka,Shokri:2020cxa}. In addition, as one moves closer the transverse/longitudinal edges of the QGP, the size of the pressure anisotropies increases at all times \cite{Martinez:2009mf,Ryblewski_2013,Strickland_2014}. Faced with this, researchers suggested to find another method to formulate hydrodynamics in a momentum-space anisotropic QGP.   Recently, there have been theoretical and phenomenological studies that try to better account for large deviations from isotropy by relaxing the assumption that the QGP is close to local isotropic thermal equilibrium. To address this issue, they introduced a framework called anisotropic hydrodynamics (aHydro) in order to describe the non-equilibrium dynamics of relativistic systems, without breaking important physics constraints such as the positivity of the one-particle distribution function~\cite{Florkowski:2010cf,Martinez:2010sc,Tinti:2013vba,Alqahtani:2017mhy,Alqahtani:2017jwl,Almaalol_2019}. 

In a prior paper \cite{Strickland_2018}, comparisons between three hydrodynamic models and exact solutions of the RTA Boltzmann equation \cite{Florkowski:2013lza,Florkowski:2013lya,Florkowski:2014sfa} were presented.  It was found that linearized viscous hydrodynamics performed more poorly than the canonical  formulation of aHydro in reproducing the exact attractor for all moments.  However, although the canonical aHydro formulation \cite{Florkowski:2010cf,Martinez:2010sc} did a reasonable job in describing moments with $m>0$, Ref.~\cite{Strickland_2018} found that it did not provide a good approximation for moments with $m=0$.  The failure of the canonical formulation was postulated to be due to the fact that the exact solutions to the RTA Boltzmann equation have an explicit two-component nature and cannot be accurately described by a single ellipsoidal form.  As a result, it would be interesting to implement aHydro with a two-component ansatz for the distribution function to see if a better description of moments with $m=0$ can be achieved.  Additionally, it would be interesting to see if this also results in a quantitative improvement for higher-order moments.

In this document, we report on our progress in obtaining improved dynamical equations for anisotropic hydrodynamics through the use of an improved ansatz for the form of the underlying aHydro distribution function which explicitly includes a free streaming contribution.  We demonstrate that with this improvement one can better reproduce exact results available in the literature for the evolution of moments of the distribution function, in particular, for moments which contain no powers of the longitudinal momentum in their definition ($m=0$ moments).  Using the resulting dynamical equations, we extract the non-equilibrium attractor associated with our improved aHydro ansatz and demonstrate that the improvement also allows one to better reproduce the exact dynamical attractor obtained using kinetic theory in the relaxation time approximation, particularly at early rescaled times and for $m=0$ moments.  We will focus our attention in this first work on a conformal system undergoing boost-invariant and transversally homogeneous Bjorken expansion, however, the method introduced herein is easily extended to full 3+1d.

The paper is organized as follows. In Sec. \ref{sec:setup} we present the basic setup and assumptions used for the system and introduce our improved aHydro distribution function ansatz.  We then use the first and second moments of the Boltzmann equation to obtain equations of motion for the dynamical parameters appearing in the new ansatz.  We do this explicitly for a system undergoing boost-invariant 0+1d Bjorken expansion.  Using the resulting dynamical equations we obtain the time evolution of all moments of the distribution function.  In Sec. \ref{sec:numerics}, we present our numerical results and discuss. In Sec. \ref{sec:conclusions} we present our conclusions and an outlook for the future.

%%%%%%%%%%%%%%%%%%%%%%%%%%%%%%%%%%%%%%%%%%%%%%%%%%%%%%%%%%%
\section{Setup}
\label{sec:setup}
%%%%%%%%%%%%%%%%%%%%%%%%%%%%%%%%%%%%%%%%%%%%%%%%%%%%%%%%%%%

For the current work, we assume a system of massless particles. Furthermore, we assume that the system is undergoing boost invariant longitudinal expansion $(v_{z}=z/t)$ and expands only along the beam-line axis, ignoring the effects of transverse dynamics.  Accordingly one can assume a homogeneous distribution in the transverse directions and set \mbox{$v_{x,y}=0$}. By taking into account these assumptions only proper-time derivatives remain and the dynamics reduces to 0+1d dimensional evolution \cite{Bjorken:1982qr}. In order to better describe free streaming contributions to the evolution of the one-particle distribution function,  we propose an improved aHydro one-particle distribution function of the form
\be
f({\bf p},\tau) = f_0(\xifs,\Lambda_0) D(\tau,\tau_0) + \frs(\xi,\Lambda) [1 - D(\tau,\tau_0)]  \, ,
\label{eq:fdef}
\ee
where the first term is the free-streaming contribution and the second term is the equilibrating contribution.  In Eq.~\eqref{eq:fdef}, $\Lambda_0$ is the initial momentum scale, $f_0$ is the initial particle distribution with
\be
\xifs = (1+\xi_0)\frac{\tau^2}{\tau_0^2} - 1 \, ,
\ee
where $\xi_0$ is the initial momentum-space anisotropy, $\tau_0$ is the initial proper time,
and 
\be
f_{\rm RS}(\xi,\Lambda) = f_{\rm eq}(\sqrt{{\bf p}^2 + \xi p_z^2}/\Lambda) \, ,
\ee
where RS indicates the anisotropic Romatschke-Strickland form \cite{Romatschke:2003ms}. The equilibrium distribution function $f_{\rm eq}$ may be taken to be a Bose-Einstein, Fermi-Dirac, or Boltzmann distribution. Here we will assume that $f_{\rm eq}$ is given by a Boltzmann distribution.  For free-streaming distribution function $f_0$ is also of RS form but with $\xi = \xifs$ and $\Lambda = \Lambda_0$, i.e.
\be
f_0(\xifs,\Lambda_0) = f_{\rm RS}(\xifs,\Lambda_0) \, .
\ee

Here we use the label `0' to emphasize that this contribution is constrained by the initial condition for the distribution function. Additionally, $ -1 < \xi < \infty $ is a parameter that indicates the strength and type of momentum-space anisotropy.  By stretching \mbox{($-1 < \xi < 0$)} or squeezing ($\xi  > 0$) the underlying isotropic distribution function $f_{\rm eq}$ along one direction in momentum-space, one can obtain an anisotropic distribution function.

In Eq.~\eqref{eq:fdef} we have also introduced the damping function $D(\tau,\tau_0)$ 
\be
D(\tau,\tau_0) = \exp\Biggr[-\int\limits_{\tau_0}^{\tau} \frac{d\tau^{\prime\prime}}{\tau_{\rm eq}(\tau^{\prime\prime})} \Biggr] , 
\label{df}
\ee
which, for finite $\tau_{\rm eq}$, obeys $\lim_{\tau \rightarrow \tau_0} D(\tau,\tau_0) = 1$ and $\lim_{\tau \rightarrow \infty} D(\tau,\tau_0) = 1$.  Note that, since $D(\tau_0,\tau_0)=1$, at $\tau=\tau_0$ the distribution function \eqref{eq:fdef} reduces to the initial distribution function $f_0$.  We note for future use that the damping function satisfies 
\be
\frac{\partial D(\tau,\tau_0)}{\partial \tau} = -\frac{D(\tau,\tau_0)}{\tau_{\rm eq}(\tau)} \, .
\label{Dprop}
\ee

\subsection{Moments of the improved distribution function}

To calculate the energy density and pressures in the local rest frame (LRF), one can integrate the distribution function \eqref{eq:fdef} times $p^\mu p^\nu$ using the Lorentz-invariant integration measure 
\be
\int dP = \int \frac{d^4{\bf p}}{(2\pi)^4} 2\pi\delta(p^2-m^2)2\theta(p^0) = \int \frac{d^3{\bf p}}{(2\pi)^3} \frac{1}{E} \, .
\ee

After performing this operation one finds that all moments of the distribution function can be decomposed into two terms, e.g.
\be
\epsilon = T^{00} = \epsilon_0(\xifs,\Lambda_0) D(\tau,\tau_0) + \epsilon_{\rm RS}(\xi,\Lambda) [1 - D(\tau,\tau_0)]  \, ,
\label{eq:epsilon1}
\ee
\be
P_L = T^{zz} = P_{L,0}(\xifs,\Lambda_0) D(\tau,\tau_0) + P_{L, \rm RS}(\xi,\Lambda) [1 - D(\tau,\tau_0)]  \, ,
\label{eq:pl1}
\ee
where in Eqs.~\eqref{eq:epsilon1} and \eqref{eq:pl1} the left hand sides are the non-equilibrium energy density and longitudinal pressure, respectively.  For a conformal system, one can use $\epsilon= 2P_T+P_L$ to determine the transverse pressure 
\be
P_T =P_{T,0}(\xifs,\Lambda_0) D(\tau,\tau_0) + P_{T, \rm RS}(\xi,\Lambda) [1 - D(\tau,\tau_0)]  \, .
\label{eq:pt1}
\ee
In general, one can compute a large set of moments of the one-particle distribution function \eqref{eq:fdef} of
the form
\be
{\cal M}^{nm} [f] = \int dP \, (p\cdot u)^n (p\cdot z)^{2m} f(\bf p) \, .
\ee
Taking a general moment of Eq.~\eqref{eq:fdef}, one finds
\be
{\cal M}^{nm}[f] = {\cal M}^{nm}[f_0] D(\tau,\tau_0) + {\cal M}^{nm}[\frs] [1 - D(\tau,\tau_0)] \, .
\ee
Note that certain moments map to familiar hydrodynamics variables, e.g. taking $n=1$ and $m=0$, one obtains the number density
\be 
n={\cal M}^{10} = \int dP \, (p\cdot u) f({\bf p}) = {\cal M}^{10}[f_0] D(\tau,\tau_0) + {\cal M}^{10}[\frs] [1 - D(\tau,\tau_0)] \, . \nonumber \\
\ee
Taking $n=2$ and $m=0$, one can evaluate the energy density via
\be
\epsilon={\cal M}^{20}= \int dP \, (p\cdot u)^2 f({\bf p}) ={\cal M}^{20}[f_0] D(\tau,\tau_0) + {\cal M}^{20}[\frs] [1 - D(\tau,\tau_0)]  \, , \nonumber \\
\ee
and taking $n=0$ and $m=1$, one obtains the longitudinal pressure 
\be
P_L ={\cal M}^{01}= \int dP \, (p\cdot z)^{2} f({\bf p}) ={\cal M}^{01}[f_0] D(\tau,\tau_0) + {\cal M}^{01}[\frs] [1 - D(\tau,\tau_0)]  \,   .\nonumber \\
\ee

Since both the free-streaming and equilibrating contributions are of RS form, we can compute the moments for both of these contributions using \cite{Strickland:2018ayk}
\be
{\cal M}_{\rm aHydro}^{nm}(\tau)=\frac{\Lambda^{2m+n+2} \Gamma (2m+n+2)}{(2 \pi )^2} {\cal H}^{nm}\left(\frac{1}{\sqrt{1+\xi}}\right)  ,
\ee
with 
\be
{\cal H}^{nm}(y)=\frac{2y^{2m+1}}{2m+1} \, {}_2F_1(\frac{1}{2}+m,\frac{1-n}{2};\frac{3}{2}+m;1-y^2) \, .
\ee
where ${}_2F_1$ is a hypergeometric function, $y = 1/\sqrt{1+\xi}$, and it has been assumed that the underlying isotropic distribution function is a Boltzmann distribution function.  In practice, we will scale these moments by their equilibrium limit, which assuming Boltzmann statistics, gives
\be
{\cal M}_{\rm eq}^{nm}(\tau) =\frac{2 T^{2 m+n+2} \Gamma (2 m+n+2)}{(2 \pi )^2 (2 m+1)} \, .
\ee
 
Using the improved aHydro ansatz \eqref{eq:fdef} one obtains
 \be
\overline{\cal M}^{nm}[f] = \frac{{\cal M}^{nm}[f_0] D(\tau,\tau_0) + {\cal M}^{nm}[\frs] [1 - D(\tau,\tau_0)]}{{\cal M}_{\rm eq}^{nm}(\tau)} \, ,   
\label{eq:mbar}
\ee
where we have introduce the scaled moments
 \be
\overline{\cal M}^{nm}(\tau) = \frac{{\cal M}^{nm}(\tau)} {{\cal M}_{\rm eq}^{nm}(\tau)} \, .
\ee
Note that one has $\overline{\cal M}_{\rm aHydro}^{nm}(\tau)$=1 if the system is in equilibrium. 
 
%%%%%%%%%%%%%%%%%%%%%%%%%%%%%%%%%%%%%%%%%%%%%%%%%%%%%%%%
\subsubsection{First moment}
%%%%%%%%%%%%%%%%%%%%%%%%%%%%%%%%%%%%%%%%%%%%%%%%%%%%%%%%

Our starting point is the Boltzmann equation for massless particles 
\be
p^\mu \partial_\mu f = C[f] \, ,
\ee
where the collisional kernel is taken to be the relaxation-time approximation (RTA) collisional kernel
\be
C[f] = - \frac{p \cdot u}{\tau_{\rm eq}(T)} [ f - f_0(T) ] \, ,
\ee
and $u^\mu$ is the four-velocity associated with the local rest frame.  Herein we will focus our attention on a system that is transversally homogenous and subject to boost-invariant Bjoken flow (0+1d). In order to preserve conformal invariance, the equilibration time must be inversely proportional to the local temperature and, for RTA, is given by \cite{ Denicol_2010, Denicol_2011} 
\be
\tau_{\rm eq}(T) = 5\bar\eta/T \, ,
\ee
where $\bar\eta = \eta/s$ is the ratio of the shear viscosity $\eta$ to entropy density $s$. 

The first moment of the left-hand side of the Boltzmann equation reduces to $\partial_\mu T^{\mu\nu}$; however, in the relaxation time approximation the first moment of the collisional kernel the right hand side results in a constraint that must be satisfied in order to conserve energy and momentum, i.e.$\int dP \, p^\mu C[f]=0$.   This constraint is referred to as the {\em matching condition} and allows one to compute the local effective temperature of the system.  In RTA, it results in the following constraint equation
\be
\epsilon_\text{eq}(\tau) = \epsilon_\text{non-eq}(\tau) \, ,
\label{eq:matchcon}
\ee
where the effective temperature $T$ appears in $\epsilon_\text{eq}$.

As a result of this constraint, computing the first moment gives
\be
\partial_\mu T^{\mu\nu}=0 \, .
\label{eq:1m}
\ee
Expanding this equation out in terms of the non-vanishing components of the energy-momentum tensor, for a 0+1d system, one obtains
\be
\dot{\epsilon} = - \frac{\epsilon + P_L}{\tau} \, .
\label{eq:eom1}
\ee
Plugging Eqs.~\eqref{eq:epsilon1} and \eqref{eq:pl1} into \eqref{eq:eom1} one finds 
\ba
&&
[1 - D(\tau,\tau_0)] \left[  \hat\Lambda^4 {\cal R}'(\xi) \dot\xi + 4 \hat\Lambda^3 {\cal R}(\xi) \dot{\hat\Lambda} + \frac{{\cal R}(\xi)  \hat\Lambda^4}{\tau} \left( 1+\frac{1}{3} \frac{{\cal R}_L(\xi)}{{\cal R}(\xi)} \right) \right] + \nonumber \\
&&
\hspace{1cm} D(\tau,\tau_0) \left[ {\cal R}'(\xifs) \dot\xi_{\rm FS} - \left( \frac{1}{\tau_{\rm eq}} - \frac{1}{\tau} \right) {\cal R}(\xifs) + \frac{1}{3\tau}  {\cal R}_L(\xifs)  + \frac{\hat\Lambda^4 {\cal R}(\xi)}{\tau_{\rm eq}}
\right] = 0 \, ,
\label{eq:d1}
\ea
with $\hat\Lambda = \Lambda/\Lambda_0$ and 
\ba
{\cal R}(\xi) &=& \frac{1}{2} \left[ \frac{1}{1+\xi} + \frac{\arctan\sqrt{\xi}}{\sqrt{\xi}}\right]   , \nonumber \\
{\cal R}_T(\xi) &=& \frac{3}{2\xi}\left[ \frac{1+(\xi^2-1){\cal R}(\xi)}{\xi+1}\right]    , \nonumber \\
{\cal R}_L(\xi) &=& \frac{3}{\xi}\left[ \frac{(\xi+1){\cal R}(\xi)-1}{\xi+1}\right]    ,
\ea
which satisfy $3 {\cal R}= 2{\cal R}_T + {\cal R}_L$. 
%\redflag{For a conformal system, the energy-momentum tensor is traceless, which implies that $\epsilon = 2P_{T} + P_{L}$.  Using this condition for any anisotropic distribution function \eqref{eq:fdef} is satisfied by Eqs.~\eqref{eq:epsilon1},\eqref{eq:pl1}, and \eqref{eq:pt1} since for an isotropic conformal case $\epsilon_{iso} = 3P_{iso}$ where $\epsilon_{iso}$ and $P_{iso}$ are the isotropic energy density and pressure, respectively.}   

%%%%%%%%%%%%%%%%%%%%%%%%%%%%%%%%%%%%%%%%%%%%%%%%%%%%%%%%
\subsubsection{Matching condition}
%%%%%%%%%%%%%%%%%%%%%%%%%%%%%%%%%%%%%%%%%%%%%%%%%%%%%%%%

At any time, we define the local effective temperature $T(\tau)$ of the fluid using the canonical matching condition which results from the vanishing of the right-hand-side of the first moment of the Boltzmann equation.  Using the improved form one finds
\be
T = {\cal R}_{\rm eff}^{1/4} \Lambda_0 \, ,
\label{eq:teff}
\ee
with
\be
{\cal R}_{\rm eff} \equiv D(\tau,\tau_0)  {\cal R}(\xifs) + [1 - D(\tau,\tau_0)] {\cal R}(\xi) \hat\Lambda^4 \, .
\ee

%%%%%%%%%%%%%%%%%%%%%%%%%%%%%%%%%%%%%%%%%%%%%%%%%%%%%%%%
\subsubsection{Second moment}
%%%%%%%%%%%%%%%%%%%%%%%%%%%%%%%%%%%%%%%%%%%%%%%%%%%%%%%%

To close the system of equations, we use the $uzz$ projection of the second-moment minus the 1/3 of the sum of $uxx$, $uyy$, and $uzz$.  That brings us to the second moment equations.  For the second moment equation of motion, we will perform a similar manipulation by starting from the relaxation-time approximation (RTA) Boltzmann equation 
\be
p^\mu \partial_\mu f = - \frac{p \cdot u}{\tau_{\rm eq}(T)} [ f - f_0(T) ] \, .
\ee
We then encounter a rank three tensor which is defined as $I^{\mu\nu\lambda}[f] \equiv N_{\rm dof} \int dP \, p^\mu p^\nu p^\lambda \, f$, where $N_{\rm dof}$ is the number of degrees of freedom.
One obtains the following equation of motion from the second moment of the RTA Boltzmann equation  \cite{Nopoush_2014}
\be
\partial_\mu I^{\mu\nu\lambda} = \frac{1}{\tau_{\rm eq}} ( u_\mu I^{\mu\nu\lambda}_{\rm eq} - u_\mu I^{\mu\nu\lambda} ) \, .
\label{eq:secondmom1}
\ee
Note that $I^{\mu\nu\lambda}$ is symmetric with respect to interchanges of $\mu$, $\nu$, and $\lambda$ and traceless in any pair of indices (massless particles/conformal invariance).

In an isotropic system, one finds $I_x = I _y = I_z = I_0$ with
\be
I_0(\Lambda) = \frac{4 N_{\rm dof}}{\pi^2} \Lambda^5 \, .
\ee
Using the canonical aHydro form one finds
\ba
I_u &=& {\cal S}_u(\xi) I_0(\Lambda) \, , \nonumber \\
I_x &=& I_y = {\cal S}_T(\xi) I_0(\Lambda) \, , \nonumber \\
I_z &=& {\cal S}_L(\xi) I_0(\Lambda) \, ,
\ea
with
\ba
{\cal S}_u(\xi) &=& \frac{3+2\xi}{(1+\xi)^{3/2}} \, , \nonumber \\
{\cal S}_T(\xi) &=& \frac{1}{\sqrt{1+\xi}} \, , \nonumber \\
{\cal S}_L(\xi) &=& \frac{1}{(1+\xi)^{3/2}} \, ,
\ea
which satisfy $2 {\cal S}_T + {\cal S}_L = {\cal S}_u$.

The $i = \{x,y,z\}$ equations result from
\be
D I_i + I_i (\theta - 2 \theta_i) = \frac{1}{\tau_{\rm eq}} ( I_{\rm eq} - I_i ) \, ,
\ee
with the co-moving derivative $D = u^\mu \partial_\mu$, the expansion scalar $\theta = \partial_\mu u^\mu$ , and $\theta_i \equiv -u_\mu D_i X_i^\mu$.
For the case of 0+1d Bjorken expansion one has $D = \partial_\tau$, $\theta = \partial_\mu u^\mu = 1/\tau$, $\theta_x = \theta_y = 0$ and $\theta_z =  - 1/\tau$.

Based on this, one has
\ba
\partial_\tau I_x + \frac{1}{\tau} I_x &=& \frac{1}{\tau_{\rm eq}} ( I_{\rm eq} - I_x ) \, , \nonumber \\
\partial_\tau I_y + \frac{1}{\tau} I_y &=& \frac{1}{\tau_{\rm eq}} ( I_{\rm eq} - I_y ) \, , \nonumber \\
\partial_\tau I_z + \frac{3}{\tau} I_z &=& \frac{1}{\tau_{\rm eq}} ( I_{\rm eq} - I_z ) \, . 
\ea

The first two equations ($xx$ and $yy$ projections) both give
\ba
&&
[1 - D(\tau,\tau_0)]\left[\frac{1}{\tau}+\frac{{\cal S}'_T(\xi)} {{\cal S}_T(\xi)} \dot\xi+\frac{5\dot{\hat\Lambda}}{\hat\Lambda}\right] + D(\tau,\tau_0) \left[  \frac{1}{\tau_{\rm eq}} +  \frac{{\cal S}'_T(\xifs)} {\hat\Lambda^5{\cal S}_T(\xi)} \xifsdot + \left( \frac{1}{\tau} - \frac{1}{\tau_{\rm eq}} \right) \frac{{\cal S}_T(\xifs)} {\hat\Lambda^5{\cal S}_T(\xi)} \right] \nonumber \\
&&
\hspace{1cm} =\frac{1}{\tau_{\rm eq}} \left[  \frac{T^5}{\Lambda_0^5 \hat\Lambda^5 {\cal S}_T(\xi)}  - D(\tau,\tau_0) \frac{{\cal S}_T(\xifs)} {\hat\Lambda^5{\cal S}_T(\xi)} - [1 - D(\tau,\tau_0)]\right] .
\label{eq:nIxxeq}
\ea
The third equation ($zz$ projection) gives
\ba
&&
[1 - D(\tau,\tau_0)]\left[\frac{3}{\tau}+\frac{{\cal S}'_L(\xi)} {{\cal S}_L(\xi)} \dot\xi+\frac{5\dot{\hat\Lambda}}{\hat\Lambda}\right] + D(\tau,\tau_0) \left[  \frac{1}{\tau_{\rm eq}} +  \frac{{\cal S}'_L(\xifs)} {\hat\Lambda^5{\cal S}_L(\xi)} \xifsdot + \left( \frac{3}{\tau} - \frac{1}{\tau_{\rm eq}} \right) \frac{{\cal S}_L(\xifs)} {\hat\Lambda^5{\cal S}_L(\xi)} \right]  \nonumber \\
&&
\hspace{1cm} =\frac{1}{\tau_{\rm eq}} \left[  \frac{T^5}{\Lambda_0^5 \hat\Lambda^5 {\cal S}_L(\xi)}  - D(\tau,\tau_0) \frac{{\cal S}_L(\xifs)} {\hat\Lambda^5{\cal S}_L(\xi)} - [1 - D(\tau,\tau_0)]\right] .
\label{eq:nIzzeq}
\ea
Taking the $zz$ projection minus one-third of the sum of the $xx$, $yy$, and $zz$ projections gives
\be
[ 1 - D(\tau,\tau_0)] \left( \frac{1}{1+\xi} \dot\xi - \frac{2}{\tau}\right) + \frac{\xi \sqrt{1+\xi}}{\tau_{\rm eq} }  \frac{\hat{T}^5}{\hat\Lambda^5}  = 0\, .
\label{eq:2ndmomfnew}
\ee
We note that all of the free streaming contributions vanish.  Solving for $\dot\xi$ using \eqref{eq:2ndmomfnew} we obtain
\be
\dot\xi =({1+\xi}) \left(\frac{2}{\tau}-\frac{\xi \sqrt{1+\xi}}{\tau_{\rm eq} } \, \frac{\hat{T}^5}{\hat\Lambda^5}\, \frac{1}{1 - D(\tau,\tau_0)}\right) .
\label{eq:2ndmomfnewxi}
\ee
As mentioned previously, in the limit $\tau \rightarrow \tau_0$, one has $D=1$ and hence the second term on the right-hand-side of \eqref{eq:2ndmomfnewxi} will diverge at $\tau = \tau_0$ unless either $\xi = 0$ or $\xi = -1$.  The latter condition makes the entire right hand side vanish and hence does not allow for dynamical evolution of $\xi$.  For this reason we will use $\lim_{\tau \rightarrow \tau_0} \xi(\tau) = 0$.

\vspace{5mm}
\paragraph*{Cross check $(D=0)$\\[0.8em]}
As cross-check on our results, one can recompute the second-moment equation with $D=0$ to see if it agrees with results available in the literature.
In this case on finds that the $zz$ projection gives
\be
(\log {\cal S}_L)' \dot\xi + 5 \partial_\tau \!\log \Lambda + \frac{3}{\tau} = \frac{1}{\tau_{\rm eq}} \left[ \frac{{\cal R}^{5/4}}{{\cal S}_L} -1 \right] ,
\label{eq:Izzeq}
\ee
and the $xx$ and $yy$ projections both give
\be
(\log {\cal S}_T)' \dot\xi + 5 \partial_\tau \!\log \Lambda + \frac{1}{\tau} = \frac{1}{\tau_{\rm eq}} \left[ \frac{{\cal R}^{5/4}}{{\cal S}_T} -1 \right] ,
\label{eq:Ixxeq}
\ee
where, in both cases, we used $T = {\cal R}^{1/4}(\xi) \Lambda$.

Finally, with $D=0$, taking the $zz$ projection minus one-third of the sum of the $xx$, $yy$, and $zz$ projections gives
\be
\frac{1}{1+\xi} \dot\xi - \frac{2}{\tau} + \frac{{\cal R}^{5/4}(\xi)}{\tau_{\rm eq}} \xi \sqrt{1+\xi} = 0\, .
\label{eq:2ndmomf}
\ee
One can verify explicitly that Eq.~\eqref{eq:2ndmomfnew} reduces to this in the limit $D \rightarrow 0$.

 %%%%%%%%%%%%%%%%%%%%%%%%%%%%%%%%%%%%%%%%%%%%%%%%%%%%%%%%
\section{Numerical solution of the dynamical equations and the anisotropic attractor}
\label{sec:numerics}
%%%%%%%%%%%%%%%%%%%%%%%%%%%%%%%%%%%%%%%%%%%%%%%%%%%%%%%%

In this section we present some representative numerical solutions using different initial conditions along with the attractor solution to which they flow.  For this purpose, we solve the first and second differential equations corresponding to Eq.~\eqref{eq:d1} and ~\eqref{eq:2ndmomfnew} for the evolution of $\xi(\tau)$ and $\Lambda(\tau)$.  However to evolve these equations we need to know the damping function.  Herein, we solve the integral equation by using an iterative method. In the first iteration, we assume that the temperature evolution contained within the integral defining $D(\tau,\tau_0)$ is given by ideal hydrodynamics, i.e. $T_{\rm guess}(\tau) = T_0 (\tau_0/\tau)^{1/3}$.  We then solve the dynamical equations \eqref{eq:d1} and ~\eqref{eq:2ndmomfnew}.  From this we obtain the approximate dependence of the effective temperature $T$ on proper time using Eq.~\eqref{eq:teff}.  The resulting effective temperature $T(\tau)$ is then used to load the damping function for the next iteration.  We repeat this process until the effective temperature and longitudinal pressure converge to a part in $10^8$.  In practice, this can be achieved with only five iterations.  Once converged, the solutions for $\xi(\tau)$ and $\Lambda(\tau)$ can be used to compute the full distribution function using Eq.~\eqref{eq:fdef} and all moments of the distribution function using Eq.~\eqref{eq:mbar}.\footnote{One can substantially reduce the number of iterations required by initializing instead with the canonical aHydro evolution equations.}

%%%%%%%%%%%%%%%%%%%%%%%%%%%%%%%%%%%%%%%%%%%%%%%%%%%%%%%%%%%
\begin{figure}[t!]
\centerline{
\includegraphics[width=0.65\linewidth]{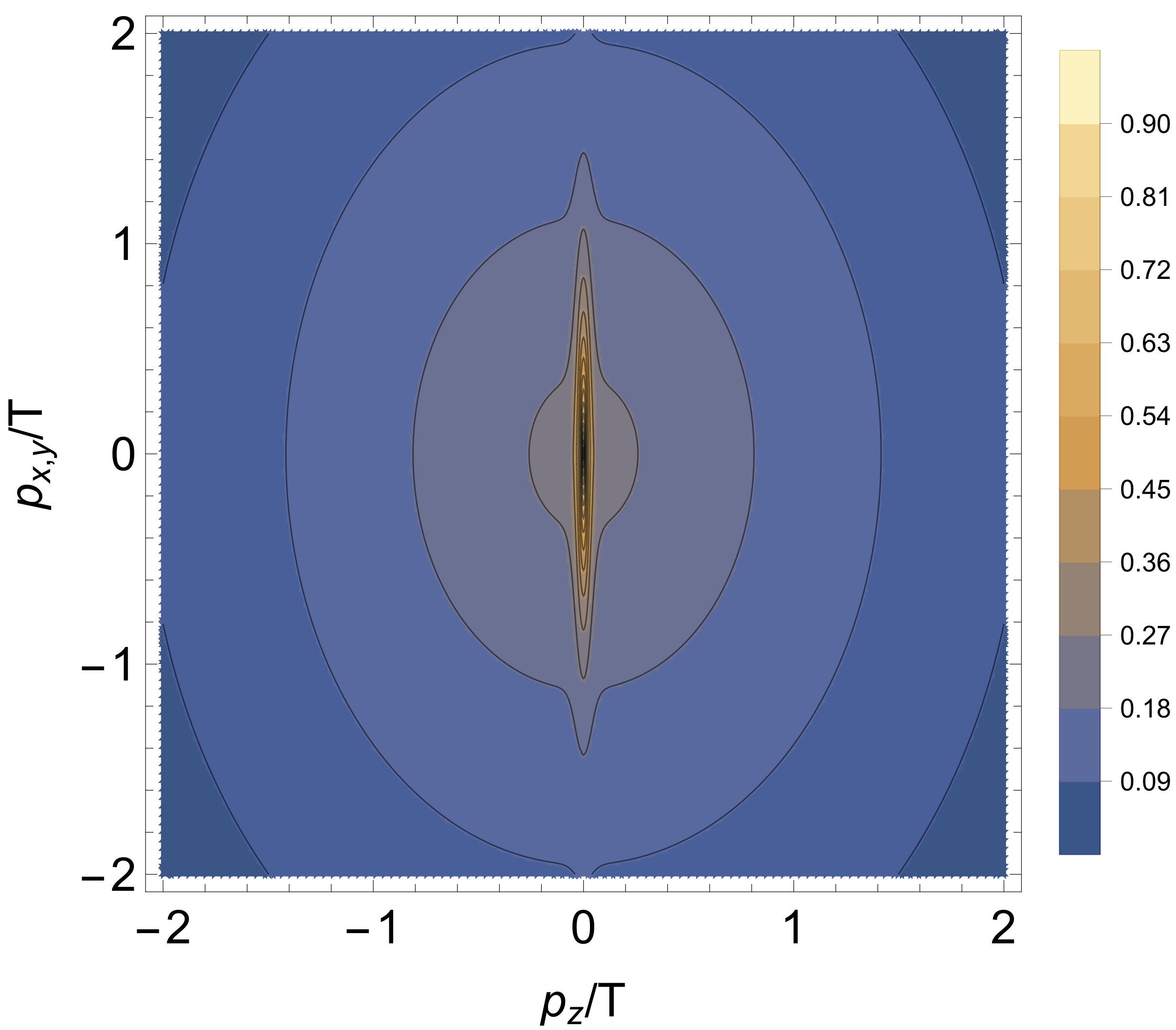}
}
\caption{Visualization of the one-particle distribution function at a given moment in proper time.  A bimodal structure can be seen, with the two contributions corresponding to a highly squeezed free-streaming component (inner ellipsoid) and a less anisotropic equilibrating contribution (outer ellipsoid). }
\label{fig:distFunc}
\end{figure}
%%%%%%%%%%%%%%%%%%%%%%%%%%%%%%%%%%%%%%%%%%%%%%%%%%%%%%%%%%%

In Fig.~\ref{fig:distFunc} we present a contour plot of the one-particle distribution function at the proper time at which the contribution from the free streaming part and equilibrating part contribute equally.\footnote{This occurs when $D(\tau,\tau_0) = 1/2$.} Generically, the exact solution for the one-particle distribution function contains two independent components \cite{Florkowski:2013lza,Florkowski:2013lya,Florkowski:2014sfa,Strickland_2018}. The first component is an anisotropic part which has been squeezed in the longitudinal direction and is exponentially damped at late times.  This contribution represents the subset particles that never had any interaction at all.  Statistically, there is always such a population of particles.  As a function of time, this contribution becomes compressed along the longitudinal direction in momentum space resulting in $P_L^{\rm FS} \rightarrow 0$ as the system evolves. This contribution comes from the first term in the exact solution Eq.~\eqref{eq:fdef} which corresponds to the free streaming contribution.  Note that, because of the damping function $D(\tau,\tau_0)$ in the first term in Eq.~\eqref{eq:fdef}, the amplitude of this very narrow ridge will decrease in time exponentially. The second visible component in Fig.~\ref{fig:distFunc}  is an isotropizing part which dominates at late times.  This contribution comes from the second term in Eq.~\eqref{eq:fdef}.

%%%%%%%%%%%%%%%%%%%%%%%%%%%%%%%%%%%%%%%%%%%%%%%%%%%%%%%%%%%
\begin{figure}[t!]
\centerline{
\includegraphics[width=1\linewidth]{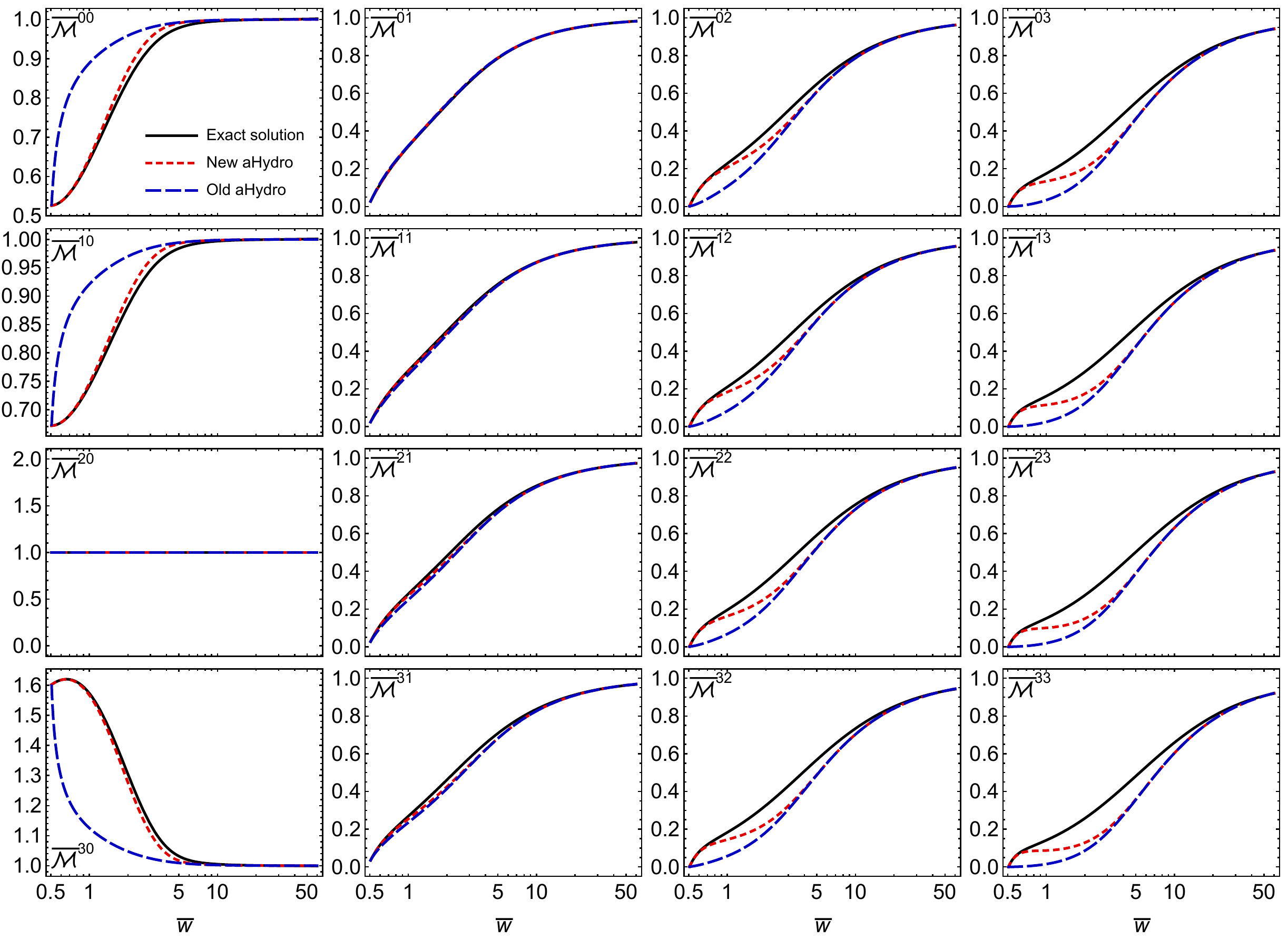}
}
\caption{Scaled moments $\overline{\cal M}^{nm}$ obtained from the exact solution (solid black line) compared with the new aHydro  (red dashed lines), and the old aHydro (blue long dashed lines). Horizontal axis is $\overline w =  {\tau} {T}/5\overline\eta$. Panels show a grid in $n$ and $m$.}
\label{fig:gridPlot}
\end{figure}
%%%%%%%%%%%%%%%%%%%%%%%%%%%%%%%%%%%%%%%%%%%%%%%%%%%%%%%%%%%

In Fig.~\ref{fig:gridPlot}, we present the evolution of the scaled moments of the distribution function as a function the scaled time 
\be
\overline w=\frac{\tau}{\tau_{eq}}=\frac{{\tau} {T}}{5\overline\eta} \, ,
\ee
and we compare to the exact RTA solution (black solid lines) obtained in Refs.~\cite{Florkowski:2013lza,Florkowski:2013lya,Strickland_2018}.  Results from the new aHydro and  old aHydro ansatze are shown as red dashed and blue dot-dashed lines, respectively. In all cases shown, the new aHydro ansatz provides a better approximation to the exact solution than the old aHydro ansatz. In addition, one observes that both aHydro ansatze result in positive definite results for all moments despite having large non-equilibrium deviations.  Comparing the old and new ansatze, we see that the new ansatz is able to reproduce the dynamics of low-order moments much better than the old ansatz.  This is particularly striking for moments with $m=0$ for which we see that the new aHydro ansatz is very close to the exact results for all $n$ shown.\footnote{We have checked that this holds true for larger $n$ than shown in Fig.~\ref{fig:gridPlot}.}  We note, however, for higher moments, e.g. $\overline{\cal M}^{33}$, we see that the new aHydro ansatz interpolates between the exact solution at early times and the old aHydro result at lates times.  As a result, one sees larger deviations from the exact solution in these moments.

\begin{figure}[t!]
\centerline{
\includegraphics[width=1\linewidth]{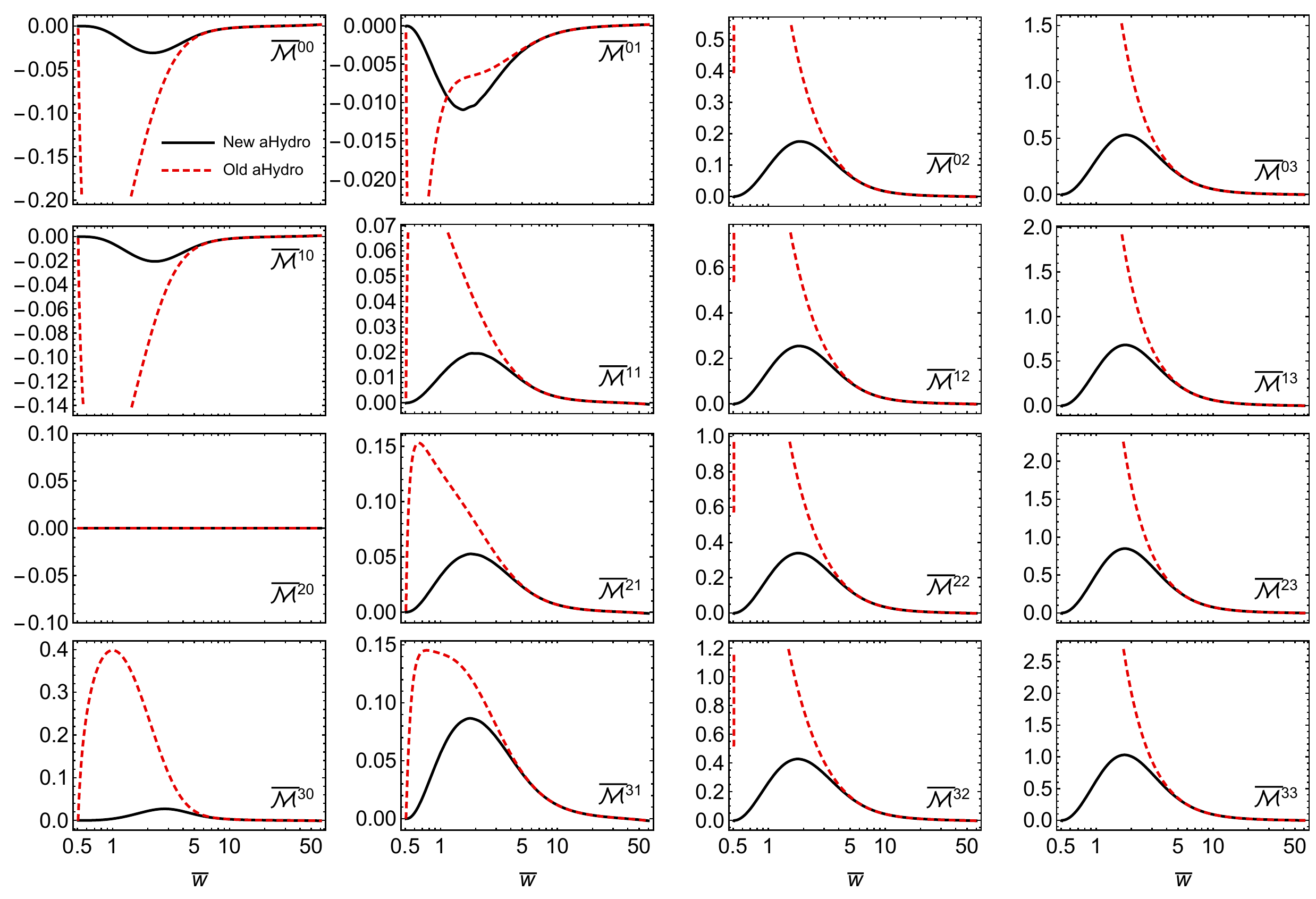}
}
\caption{Plots of the relative error between the new (solid black line) and old (red dashed lines) aHydro ansatz compared to the exact solution.  Error is computed as approximation/exact -1.}
\label{fig:errorPlot}
\end{figure}

In order to provide more quantitative comparison of the two methods, in Fig.~\ref{fig:errorPlot} we present the relative errors of the old and new aHydro ansatze computed as the ratio of a given approximation to the the exact result minus one.  The relative errors for the old and new schemes are shown at solid black and dashed red lines, respectively.  As one can see from Fig.~\ref{fig:errorPlot}, the new aHydro has a smaller error in all moments and at virtually all times.  The one exception is $\overline{\cal M}^{01}$ for which one observes a slight smaller error with the old ansatze in a small time window.  Returning to the general case, we see that, since the new scheme merges onto the old scheme at late times, they have similar relative errors, however at early time we see a dramatic reduction in the relative error using the new aHydro ansatz.
 
%%%%%%%%%%%%%%%%%%%%%%%%%%%%%%%%%%%%%%%%%%%%%%%%%%%%%%%%%%%
%%%%%%%%%%%%%%%%%%%%%%%%%%%%%%%%%%%%%%%%%%%%%%%%%%%%%%%%%%%
 
 %%%%%%%%%%%%%%%%%%%%%%%%%%%%%%%%%%%%%%%%%%%%%%%%%%%%%%%%%%%
\begin{figure}[t!]
\centerline{
\includegraphics[width=1\linewidth]{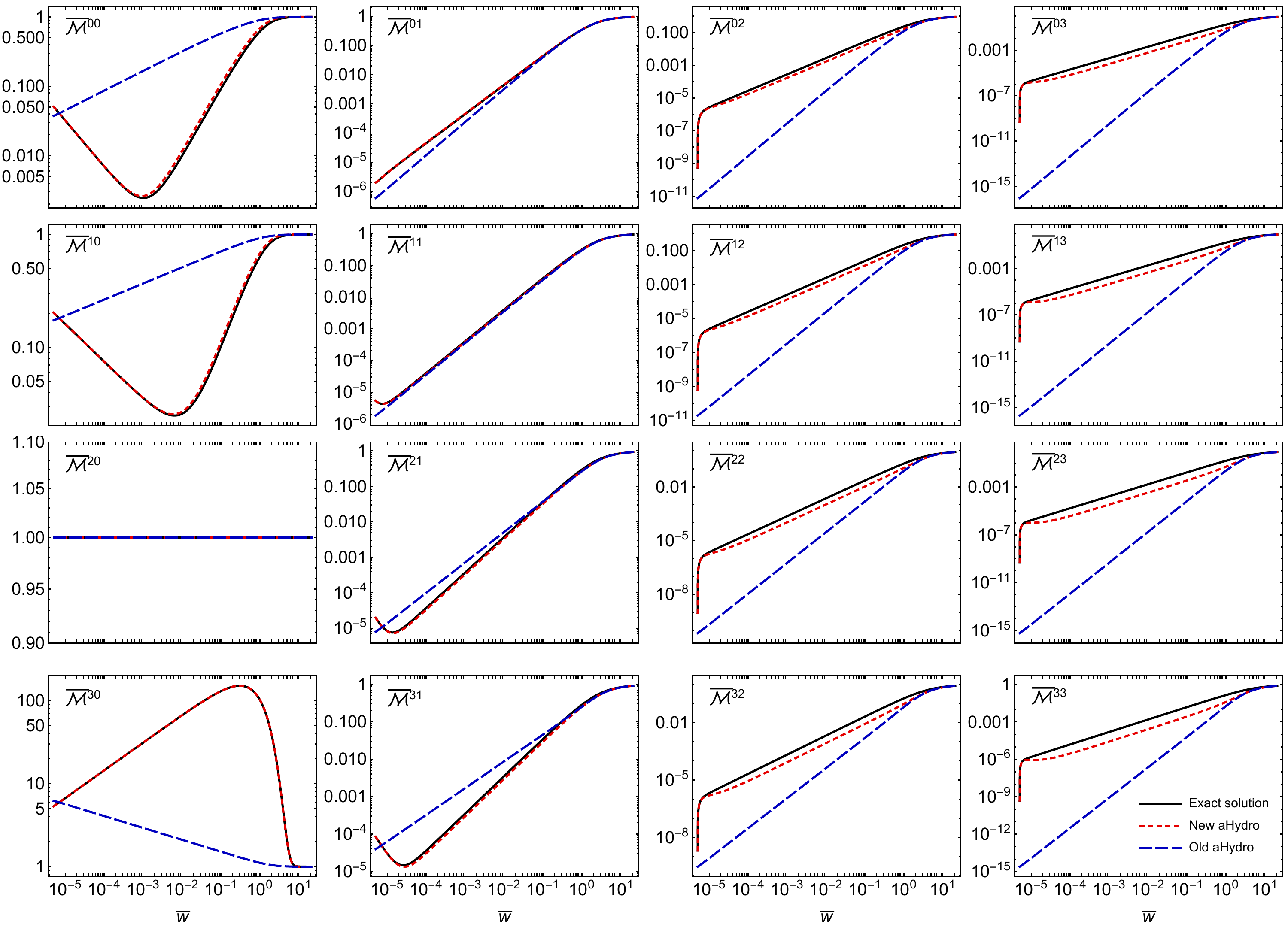}
}
\caption{Scaled moments $\overline{\cal M}^{nm}$ obtained from the exact solution attractor (solid black line) compared with the new aHydro (red dashed lines), and the old aHydro (blue long dashed lines). Horizontal axis is $\overline w$=$\tau T/5\overline\eta$. Panels show a grid in $n$ and $m$.}
\label{fig:attractorPlot}
\end{figure}
%%%%%%%%%%%%%%%%%%%%%%%%%%%%%%%%%%%%%%%%%%%%%%%%%%%%%%%%%%%

In Fig.~\ref{fig:attractorPlot}, the new aHydro (red short-dashed), and the old aHydro (blue long-dashed) attractors are compared to the attractor obtained via exact solution of the RTA Boltzmann equation (black solid line). In all cases shown, the new aHydro ansatz agrees best with the exact solution for the 0+1d conformal RTA attractor. Additionally, for all values of $\overline w$, we note that both aHydro attractors possess positive values for all moments. In the case of the new aHydro ansatz, firstly one sees that for $m=0$ and $m=1$ (first and second left column, respectively of Fig.~\ref{fig:attractorPlot})  this scheme has the best agreement at all times. As a result, the new aHydro accurately describes the evolution of the modes with $m=0$, and 1, which are sensitive to the free-streaming part of the evolution. For $m > 1$ one sees that, as $m$ and $n$ are increased, the new aHydro results differ more from the exact solutions in the region $\overline w\sim [10^{-5},1]$.   The worst agreement is for the $m = 3$ moments (rightmost column of Fig.~\ref{fig:attractorPlot}). One finds that the new aHydro ansatz fails to accurately describe the evolution of the scaled moments with $m = 3$ which are dominated by isotropizing contribution at late times.  As a consequence, the new aHydro does not provide reliable approximations for these moments and the problem becomes more severe as one increases for $m>1$.  Turning to the old aHydro ansatz,  for $m =1$, one sees that, although the old aHydro ansatz does a reasonable job in describing the $m=1$ moments, as $n$ and $m$ are increased or decreased, the results become significantly worse.  Note that, even given the caveats mentioned above, comparing the old and new aHydro ansatze, we see that the new approach dramatically improves agreement with the exact RTA attractor and, in particular, can be used to fix the problem encountered with moments with $m=0$.

 %%%%%%%%%%%%%%%%%%%%%%%%%%%%%%%%%%%%%%%%%%%%%%%%%%%%%%%%
\section{Conclusions}
\label{sec:conclusions}
%%%%%%%%%%%%%%%%%%%%%%%%%%%%%%%%%%%%%%%%%%%%%%%%%%%%%%%%

In this study, our goal was to find an improved set of anisotropic hydrodynamic evolution equations that can more faithfully describe the non-equilibrium dynamics of the quark-gluon plasma created in relativistic heavy ion collisions at RHIC and LHC. We introduced a new version of anisotropic hydrodynamics that includes separate free-streaming and equilibrating contributions which allows for a better description of exact solutions to the Boltzmann equation available in the literature. We computed explicit expressions for the first and second moments of the one-particle distribution function in the new aHydro approach and used these to obtain the new 0+1d conformal equations of motion given by Eqs.~\eqref{eq:d1} and ~\eqref{eq:2ndmomfnew}.  We presented comparisons of the numerical solution of the conformal 0+1d equations of motion for both the old and new aHydro schemes with the exact RTA solution.  Our results demonstrated that the new aHydro form allows one to have a bimodal distribution function similar to what is seen in the exact RTA solution for the one-particle distribution function.  We then computed the evolution of the scaled moments as a function the scaled time, $\overline w$, and demonstrated that the new aHydro ansatz provides a better approximation to the exact solution than the original aHydro ansatz.   Finally, we determined the non-equilibrium attractor associated with the new aHydro scheme and demonstrated that it provides much better agreement with the exact RTA attractor than the original aHydro scheme, in particular for moments with $m=0$.  In the future, it would be interesting to apply the ansatz obtained here to full 3+1d anisotropic hydrodynamics, including temperature-dependent masses for the particles similar to `canonical' quasiparticle aHydro~\cite{Alqahtani:2015qja,Alqahtani:2017mhy}.

\vspace{2mm}

\acknowledgements{
H.A. and M.S. were supported by the U.S. Department of Energy, Office of Science, Office of Nuclear Physics under Award No. DE-SC0013470.
H.A. was also supported by a visiting Ph.D. scholarship from Umm Al-Qura University.}

%%%%%%%%%%%%%%%%%%%%%%%%%%%%%%%%%%%%%%%%%%%%%%%%%%%%%%%%%%%
\bibliography{fsahydro}
%%%%%%%%%%%%%%%%%%%%%%%%%%%%%%%%%%%%%%%%%%%%%%%%%%%%%%%%%%%

%%%%%%%%%%%%%%%%%%%%%%%%%%%%%%%%%%%%%%%%%%%%%%%%%%%%%%%%%%%
% end document
%%%%%%%%%%%%%%%%%%%%%%%%%%%%%%%%%%%%%%%%%%%%%%%%%%%%%%%%%%%

\end{document}